\documentclass[a4paper]{article}

\usepackage{cite}
\usepackage{amssymb}
\usepackage{amsmath}
\usepackage{bm}

\usepackage{graphicx}

\begin{document}

\title{Weak Measurement and its Experimental Realisation.}
\author{ R. Flack$^1$ and B. J. Hiley$^{1,2}$\footnote{E-mail addresses r.flack@ucl.ac.uk; b.hiley@bbk.ac.uk.}}
\date{1.HEP, University College, Gower Street, London\\\vspace{0.4cm}2.TPRU, Birkbeck, University of London, Malet Street, \\London WC1E 7HX.}

\maketitle

\begin{abstract}

The relationship between the real part of the weak value of the momentum operator at a post selected position  is discussed and  the meaning of the experimentally determined stream-lines in the Toronto experiment of Kocsis {\em et al} is re-examined.  We argue against interpreting the energy flow lines as photon trajectories.  The possibility of performing an analogous experiment using atoms is proposed in order that a direct comparison can be made with the trajectories calculated by Philippidis, Dewdney and Hiley using the Bohm approach.

\end{abstract}

\section{Introduction}

The notion of a weak measurement has opened up new ways to explore quantum processes \cite{asaf13, aav88, dss89, kbrs11, rsh91}.  In contrast to the usual von Neumann or strong measurement, which gives information about the eigenvalues of a dynamical operator, weak measurements enable us to obtain information about small induced phase changes. This process allows us to experimentally investigate new features of a quantum process.   Weak values of the momentum operator gives us access to the components of the energy-momentum tensor which, in turn, have a direct significance for the Bohm approach.  

In fact the real part of these weak values  enable us to measure components of the energy-momentum tensor of, not only the electromagnetic field, but also the Schr\"{o}dinger, Pauli and Dirac fields \cite{bh12, rl05, hw07}.  It is here that we make contact with the Bohm approach because these components  are directly related to Bohm momentum, $\bm p_B=\nabla S$, a parameter that plays a fundamental role in this approach.

\begin{figure}[h] %  figure placement: here, top, bottom, or page
   \centering
   \includegraphics[width=2in]{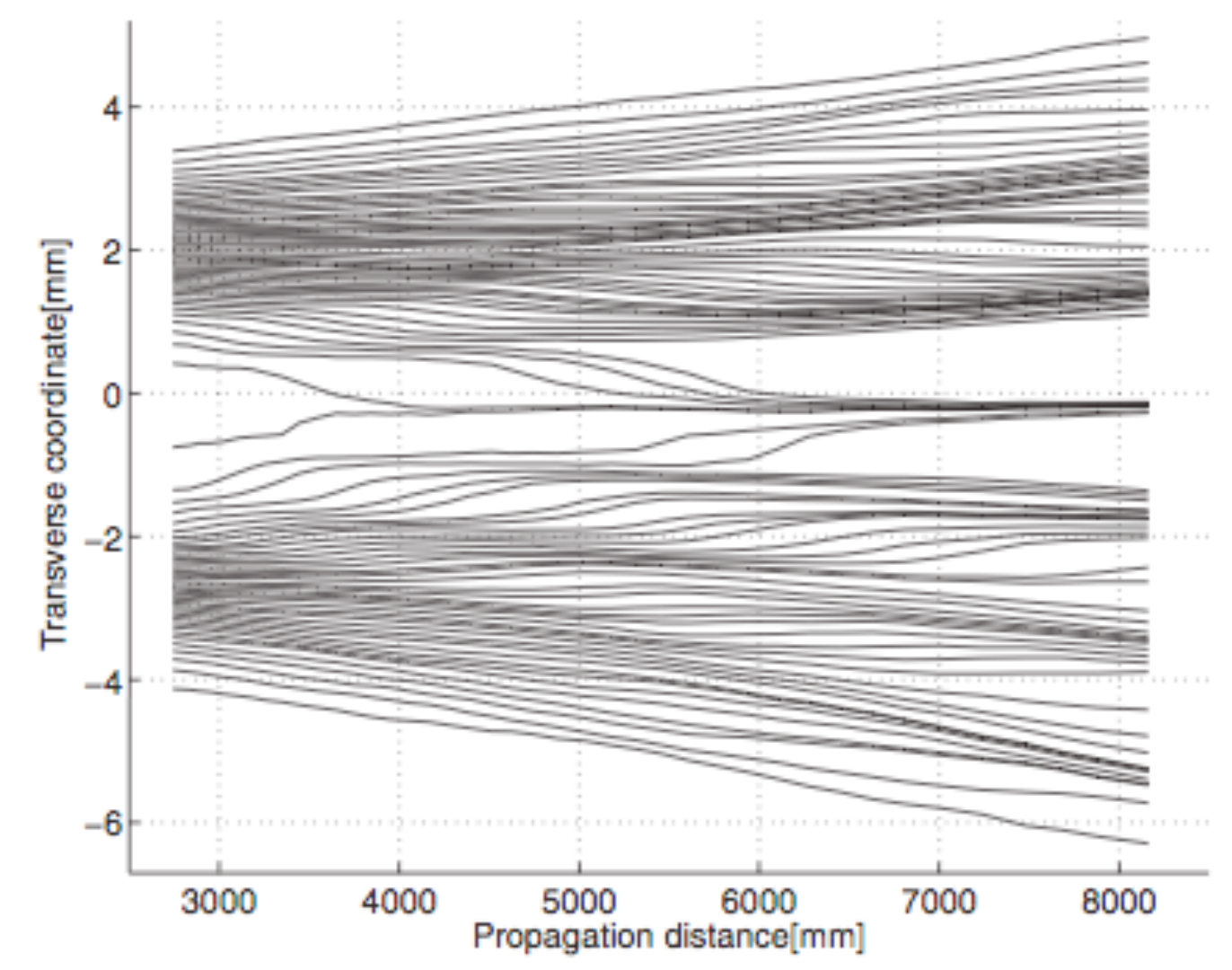} 
   \caption{Experimentally  Constructed Electromagnetic Energy Flow Lines}
   \label{fig:emflow}
\end{figure}

This connection has already been recognised and used in the experiments of Kocsis {\em et al} \cite{kbrs11}.  Here weak measurements have been made to obtain values for the Poynting vector and from these measurements, sets of energy flow lines have been constructed.   This specific experiment used a weak electromagnetic source so that only one photon entered the apparatus at a time.  The information about the weak value of the momentum operator was then found using standard photon counting techniques.  As single photons are involved, the question as to the meaning of Poynting's vector for a single photon is raised,  a question that has remained unanswered since the inception of the notion of a photon \cite{crak08}.  Can the flow lines in this experiment shown in Figure~\ref{fig:emflow} be regarded as photon trajectories?

\begin{figure}[h] %  figure placement: here, top, bottom, or page
   \centering
   \includegraphics[width=2in]{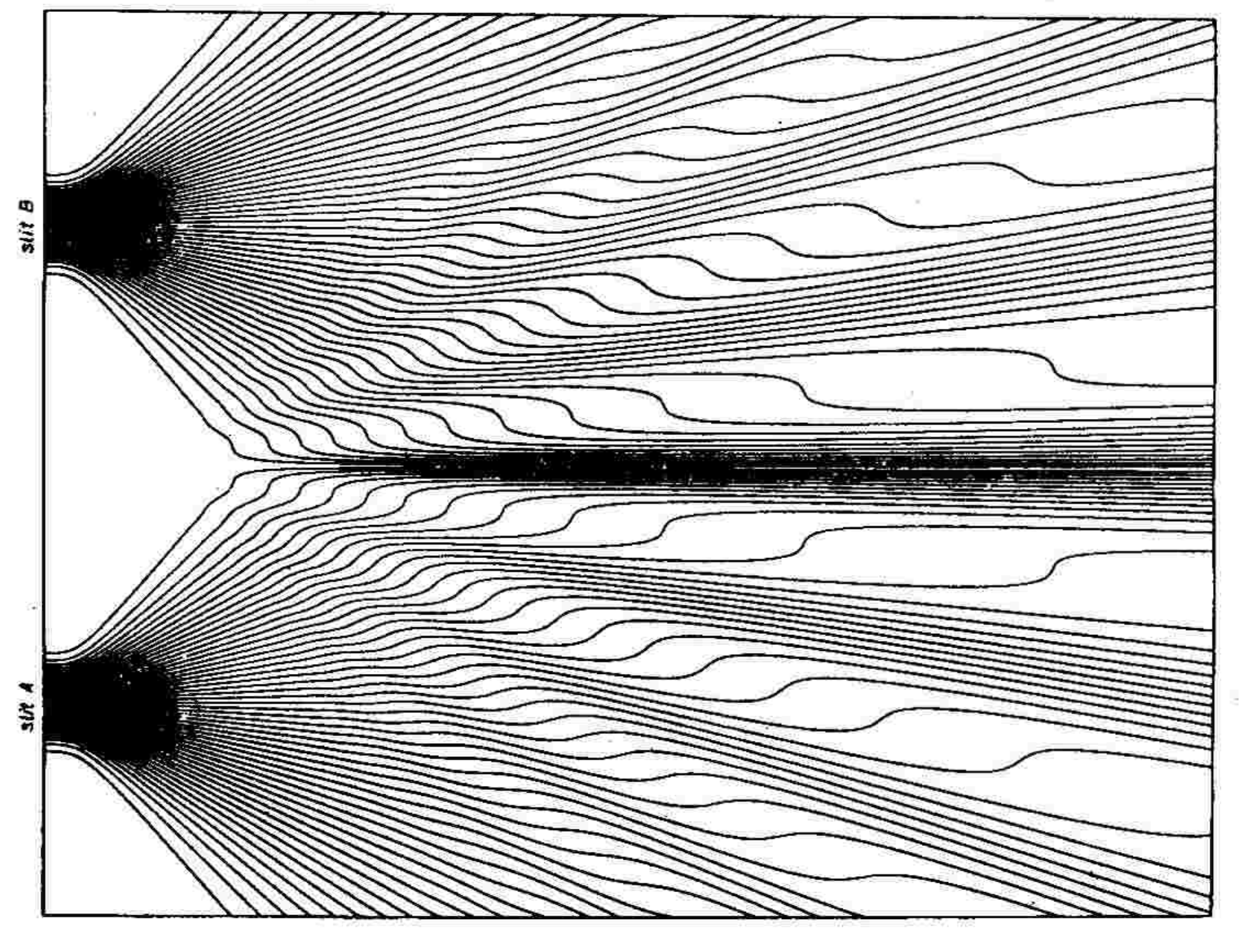} 
   \caption{Two-Slit Trajectories calculated in Bohm Approach \cite{cpcdbh79}.}
   \label{fig:traj1}
\end{figure}

The one model that could possibly give an answer to this question is that introduced by Bohm \cite{db52}.  Recall that in that model, the close connection between the classical Hamilton-Jacobi equation and the real part of the Schr\"{o}dinger equation under polar decomposition of the wave function, enables a  meaning to be given to the notion of a trajectory of a particle even in  quantum mechanical interference situations such as the two-slit experiment \cite{db52, db52a, dbbh87, dbbh93}.  The trajectories for a Schr\"{o}dinger particle are shown in Figure \ref{fig:traj1}.    Not only do these results hold for the Schr\"{o}dinger particle, they also hold for the Pauli and Dirac  particles \cite{dhkv88, bhbc12}.  In all these cases a meaning can be given to a trajectory because in a suitable non-relativistic limit, the trajectory actually becomes identical to the classical trajectory.

Although the Bohm approach allows for  the notion of a trajectory for  Schr\"{o}dinger particles, it does not give a meaning to the trajectory of a photon.  This feature has long been known and has already been discussed by Bohm \cite{db52a}, while a more comprehensive treatment can be found in Bohm, Hiley and Kaloyerou \cite{dbbhpk87} and Kaloyerou \cite{pk94}.  

In this paper we will discuss the difficulties that arise in attempting to give a meaning to the notion of  a photon trajectory.  This will lead us to suggest that  atoms should be used instead of photons.  We will also discuss the principles behind the design of an experiment that uses atoms, rather than photons, in a typical two-slit situation.  This will require us to make a weak measurement on the atoms after they have passed through the two-slits.  From these results we will be able to construct trajectories using techniques that were pioneered by  Kocsis {\em et al} \cite{kbrs11}.  In this way we will be able to compare these experimentally constructed trajectories with those calculated by Philippidis {\em et al} \cite{cpcdbh79}.  In this case we will be comparing like-with-like.

\section{What is a Weak Value?}

\subsection{Definition of Weak Value.}
We start with a formal definition of a weak value given in Aharonov {\em et al} \cite{aav88}.  The weak value of an operator $\hat A$ is 
\begin{eqnarray*}
\langle \hat A\rangle_{W_{\phi,\psi}}=\frac{\langle \phi|\hat A|\psi\rangle}{\langle\phi|\psi\rangle}
\end{eqnarray*}
The first thing we notice is that it is a complex number so it actually gives rise to two values.  Notice further that as the real part of $\langle\phi|\psi\rangle$ approaches zero, such values becomes very large, which must already throw up doubts as to the usefulness of the concept of a weak value.  However such doubts should be put aside and the idea pursued further.  Let us now turn to see what extra information the weak value of the momentum operator, $\hat P$, gives us.

\subsection{Weak Value of the Momentum Operator}

Clearly the weak value depends upon the post selected state $\phi$, so let us consider the weak value of the momentum operator at a post selected position so that we can write  $|\phi\rangle =|x\rangle$, then
\begin{eqnarray*}
\langle x|\hat P|\psi(t)\rangle=-i\nabla_x\psi(x,t)
\end{eqnarray*}
If we now write $\psi(x,t)=R(x,t)e^{iS(x,t)}$ we find
\begin{eqnarray*}
\langle \hat P\rangle_{W_{\psi(x,t)}}=\nabla S(x,t)-i\nabla_x\rho(x,t)/2\rho(x,t)
\end{eqnarray*}
One immediately recognises that the real part is the Bohm momentum, $\bm P_B=\bm\nabla S$, used in the Bohm approach to  quantum mechanics, while the imaginary part is the osmotic momentum, with which we will not be concerned in this paper.  (For details of the osmotic momentum see Hiley \cite{bh12}, Bohm and Hiley \cite{dbbh89} and references therein.)

This result may seem model specific and has little to add to standard quantum mechanics, but it should be noted that the Bohm momentum is intimately related to the $(0,j)$ component of the energy-momentum tensor, $T^{\mu\nu}$, when one treats the wave function as a field \cite{tt53, bhbc11}. In fact we find
\begin{eqnarray}
\rho P^i_B(x,t)=T^{0i}(x,t)\quad\mbox{and}\quad\rho E_B(x,t)=T^{00}(x,t).  \label{eq:BP}
\end{eqnarray}
Here $\bm P_B$ and $E_B$ are the Bohm momentum and Bohm energy respectively.  

There are two points to emphasise here.  Firstly the weak value of the momentum is the $T^{0j}$ component of the energy-momentum tensor. Secondly, it gives the possibility of attaching a value to the Bohm momentum, thus making the approach open to experimental investigation.  Furthermore it should be noted that $\bm P_B$ is not the eigenvalue momentum revealed in a strong or von Neumann measurement of the momentum operator.   This difference has been discussed at length in Belinfante \cite{fb73}.  These results are not confined to the Schr\"{o}dinger particle but also hold for the Pauli and Dirac particles \cite{bhbc11, bhbc11a}.  Thus we can now obtain additional empirical information about quantum processes.

Note further that the Bohm momentum has nothing to do with classical physics as seems to have taken root from the way Bohm originally introduced his ideas \cite{db52, db52a}.  It is actually a feature of  the quantum formalism itself.   It should also be pointed out that exactly the same results can be obtained for the Schr\"{o}dinger and Pauli  particles using the von Neumann-Moyal algebraic approach \cite{bh12}.

With this particular example we see that the weak value of the four-momentum operator can give us information about the energy-momentum content of the quantum process.  The next question is to ask how we can access these weak values experimentally.

\subsection{Weak Measurement}

The general method for extracting information in a weak measurement using the Stern-Gerlach approach has been discussed in detail by Duck {\em et al} \cite{dss89} and by Flack and Hiley \cite{rfbh14}.  In that discussion it is the weak value of the of the spin operator that is measured.  In this paper we will be interested in    the weak value of the momentum operator, so we will not discuss this example further here.   

To measure the weak value of the momentum operator  we have to find  some way of coupling the momentum operator  to some other observable which can then be used to extract the weak value $\langle P\rangle_W$.  This can be done in the final stage of the measurement where a strong, von Neumann measurement is made of this operator. This will involve a collapse of a part of the wave function without destroying the weak value.  Let us call this operator, $\hat A$, the nature of which will be specified later.  Then the interaction Hamiltonian can be written as $H_I=g(t)\hat P.\hat A$.

To see how this works, let us start with a particle in an initial state $|\Psi(t_0)\rangle=|\psi(t_0)\rangle|\xi\rangle$ where the momentum operator, $\hat P$ acts on $|\psi(t_0)\rangle$ and the  operator $\hat A$ acts on  $|\xi\rangle$.  Now introduce the time development operator $U(t,t_0)=\exp\left[-i\int ^t_{t_0} H_I(t')dt'\right]$ so that the solution of the Schr\"{o}dinger equation can be written in the form
\begin{eqnarray*}
|\Psi(t)\rangle|\xi\rangle=\exp\left[-iD\Delta t\hat P.\hat A\right]|\Psi(t_0)\rangle|\xi\rangle
\end{eqnarray*}
where we have writen
\begin{eqnarray*}
\eta=\int ^t_{t_0}g(t')dt' = D \Delta t 
\end{eqnarray*}
Here $D$ is the strength of the interaction and $\Delta t$ is the time the interaction is active. Let us denote the complete orthonormal set of eigenfunctions of $\hat A$ by $|\xi_n\rangle$, so that
\begin{eqnarray*}
\hat A|\xi_n\rangle=a_n|\xi_n\rangle
\end{eqnarray*}
Since  we will be interested in the weak value of the momentum at a specific point $x$, we will multiply from the left by $\langle x|$ and expand the $n$-th component of the  exponential to form
\begin{eqnarray*}
\langle x|e^{-i\eta a_n \hat P}|\psi(t_0)\rangle|\xi_n\rangle=\langle x|\psi(t_0)\rangle\sum_{m=0}^{m=\infty}
\frac{(-i\eta a_n)^m}{m!}\frac{\langle x|\hat P^m|\psi(t_0)\rangle}{\psi(x,t_0)}|\xi_n\rangle
\end{eqnarray*}
This can be rewritten as
\begin{eqnarray*}
\langle x|e^{-i\eta a_n\hat P}|\psi(t_0)\rangle|\xi_n\rangle=\hspace{8.5cm}\\
\hspace{1cm}=\psi(x, t_0)\left[e^{-i\eta a_n}\langle P(x)\rangle_W
+\sum _{m=2}^{\infty}\frac{(-i\eta a_n)^m}{m!}\left[\langle P^m(x)\rangle_W-\langle P(x)\rangle_W^m\right]
\right]|\xi_n\rangle
\end{eqnarray*}
  When the term involving the sum $\sum_{j=2}$  is small so that it can be neglected, we can write the final state of the particle in the form
\begin{eqnarray}
\psi(x,t)|\xi\rangle=\sum_n \psi(x,t_0)c_n e^{-i\eta a_n\langle P(x)\rangle_W}|\xi_n\rangle.  \label{eq:final state}
\end{eqnarray}
We have chosen $|\xi\rangle=\sum_nc_n|\xi_n\rangle$.  
In this way we see that the weak measurement has changed the state $|\xi\rangle$ so that it becomes
\begin{eqnarray*}
|\xi_f\rangle=\sum_nd_n|\xi_n\rangle\quad\mbox{with}\quad d_n=c_ne^{-i\eta a_n\langle P\rangle_W}.
\end{eqnarray*}
 Thus the weak value $\langle P(x)\rangle_W$ can be found from the coefficients $d_n$. This means that if we now make a suitable strong measurement on the final state $|\xi_f\rangle$, we can pick out one of the eigenstates $|\xi_n\rangle$, say,  and then from $d_n$ we can find the weak value $\langle P(x)\rangle_W$.  

If we try to determine $d_n$ simply from the probability of finding $|\xi_n\rangle$, we will not find $\langle P(x)\rangle_W$ because it appears in the imaginary exponent.  To abstract the information from these coefficients, we need to make a strong measurement on an operator complementary to $\hat A$, i.e. one that does not commute with with $\hat A$.  Let us choose $B|\mu_n\rangle= m_n|\mu_n\rangle$ with $[\hat A,\hat B]\ne 0$.  Now suppose
\begin{eqnarray*}
|\xi_j\rangle=\sum_i b_{ji}|\mu_i\rangle.
\end{eqnarray*}
then
\begin{eqnarray*}
|\xi_f\rangle=\sum_{n,i}d_nb_{in}|\mu_i\rangle
\end{eqnarray*}
so that
\begin{eqnarray*}
\hat B|\xi_f\rangle=\sum_{n,k}m_kd_nb_{kn}|\mu_k\rangle
\end{eqnarray*}
If we now make a strong measurement of $\hat B$, and find the probability of finding one particular state, say $|\mu_r\rangle$.  This probability is given by $|\sum_nd_nb_{rn}|^2$, from which we can abstract the weak value $\langle P(x)\rangle_W$ as required.

\subsection{What can Couple to the Momentum of an Atom?}

In order to find what couples to the momentum in an atomic beam, we need to remember that as  an electrically neutral atom with a magnetic dipole moment, $\bm \mu$, moves, it appears to have an  effective {\em electric} dipole moment \cite{sajzh91},
\begin{eqnarray*}
\bm d=(\bm{v\times \mu})/c
\end{eqnarray*}
The energy of this dipole in an electric field will be
\begin{eqnarray*}
U= -\bm{d.E}=\bm{v.(E\times\mu})/2
\end{eqnarray*}
so the term that we couple to the momentum operator is $(\bm{E\times \mu)}$.  If we replace the magnetic moment by its operator equivalent, $\bm\hat\mu= \frac{e\hbar}{2mc}\bm{\hat\sigma}$, our interaction Hamiltonian becomes
\begin{eqnarray}
H_I=-\frac{e\hbar}{4m^2c}(\bm{\hat P.(E\times \hat\sigma)}).   \label{eq:HI}
\end{eqnarray}
This interaction Hamiltonian has been used to demonstrate the Aharonov-Casher effect \cite{ac84, cok89}.  

We now have a specific operator, $\bm{(E\times \hat\sigma)}$  to replace the operator $\hat A$ used in Section 2.2 so that it is now possible to measure, among other things, the Bohm momentum, $\bm P_B(x)$ in various situations.  In this paper we are interested in exploring the two-slit experiment using Schr\"{o}dinger particles, namely atoms.

\section{Measurements to find Trajectories in a Two-slit Interferometer.}

\subsection{Bohm Trajectories.}

In order to proceed we first discuss exactly how the notion of a particle trajectory arises the Bohm approach.  The simplest way to see this is to follow Bohm \cite{dbbh93, db52} and  split the Schr\"{o}dinger equation into its real and imaginary parts under polar decomposition, $\psi=R\exp(iS)$\footnote{We put $\hbar=1$ throughout.}.  It is then straight forward  to show that the real part of the Schr\"{o}dinger equation becomes
\begin{eqnarray}
\partial S/\partial t +( \nabla S)^2/2m +Q+V=0.   \label{eq:HJ}
\end{eqnarray}
This equation is known as the quantum Hamilton-Jacobi equation.  We emphasise that this equation follows exactly from the Schr\"{o}dinger equation and no new mathematical structure is added.  

The equation has a remarkable similarity to classical Hamilton-Jacobi equation 
where $S$, the phase of the wave function, replaces the classical action and a new quality of energy, the quantum potential energy, $Q=-\nabla^2R/2mR$, appears.  If we assume the canonical relation $\bm P_B=\bm {\nabla S}$, is still valid  in the quantum case, then we have defined at every point in space, a momentum $\bm P_B(x)$.  
  
In contrast to the observable momentum eigenvalue, Bohm  assumed  that $\bm {P_B}$ was the actual momentum `possessed' by the particle, the ``beable", a term introduced by Bell \cite{jb87}.  In this way the particle is assumed to actually possess simultaneously a position and a momentum, where, to repeat, this momentum is not the eigenvalue of the momentum operator when we are in the $x$-representation.  Note that if we are working in the $p$-representation, then $\bm{P_B}=\bm p$.
  
 One of the original objections to the Bohm approach was that since $\bm{P_B}$ was not an observable,  it was not a meaningful quantity.  Since there seemed to be no way to attribute an experimental value to $\bm{P_B}$, it, along with the whole approach, could easily be dismissed.  However several factors have arisen over the last ten years that has clarified its meaning.  These factors taken together with the fact that the momentum corresponds to a weak value, it is now possible to explore its predictions experimentally.

The first of these factors, as has already been pointed out, is that there is a connection between the terms that are used in the Bohm model and the components of the energy momentum tensor when the Schr\"{o}dinger wave is treated as a quantum field in its own right.  In fact this relationship has already been pointed out by Takabayasi \cite{tt52} many years ago. Since then this relationship has been extended to apply to the Pauli and Dirac particles by Hiley and Callaghan \cite{bhbc11}.   It is from this work the the relationship shown in equation~(\ref{eq:BP}) was established.  Takabayasi \cite{tt52} has also shown that the quantum potential appears in the $T^{jj}$ components of the energy-momentum tensor so even the quantum potential is open to experimental determination.

Secondly the Bohm momentum is identical to a momentum defined by Moyal \cite{jm49}, the Moyal momentum.   Moyal  shows that the transport equation for this momentum is the real part of the Schr\"{o}dinger equation, which, of course, includes the quantum potential.   All of this might be dismissed as a compelling factor because ``it is simply replacing one semi-classical theory for another".       However this conclusion is not correct.  Hiley \cite{bh12} has shown that the Moyal algebra is isomorphic to the algebra introduced by von Neumann in his classic 1931 paper \cite{vn31}.   This algebra leads to the uniqueness of the Schr\"{o}dinger representation and is therefore at the heart of the quantum formalism.
These results show that the Moyal theory and, in consequence the Bohm approach, are central to standard quantum formalism. 

 What the Moyal algebra describes is a non-commutative statistical theory with the Moyal (Bohm) momentum emerging as the conditional expectation value of the momentum derived using the distribution function $F_\psi(X,P)$.  This distribution function is, in fact, a two-point density matrix, $\rho(X,P)$, where $(X,P)$ are the mean position and mean momentum of an extended region of phase space, the quantum blob in the language of de Gosson \cite{mdg12}.  

Thirdly, there is the relatively recent discovery to which we have given prominence in this paper, namely, that the Bohm momentum is simply the real part of the  weak value of the momentum operator evaluated at a post-selected position \cite{rl05, hw07, bh12}. Thus if we can measure $\langle \bm P\rangle_W$ in, say,  a two-slit experiment, we will a value of $\bm P_B(x)$ defined at all points in space from which we can construct the trajectories shown in Figure~\ref{fig:traj1}.  Furthermore if we can also find the weak value corresponding to the Bohm kinetic energy, viz. $\langle P^2\rangle_W$, then we can also put a value to the quantum potential energy $Q$, so that {\em all} the dynamic variables used in the Bohm approach can be investigated experimentally.

\subsection{Experimental Realisation using Photons.}

A preliminary investigation of the possibility of experimentally determining the form of trajectories  in a two-slit type interference apparatus has already been carried out using photons \cite{kbrs11},  but there are some difficulties in interpreting their energy flow lines as photon trajectories.  What this important experiment  actually shows is that it is possible to construct energy flow lines in this case.   

\begin{figure}[h] %  figure placement: here, top, bottom, or page
   \centering
   \includegraphics[width=2in]{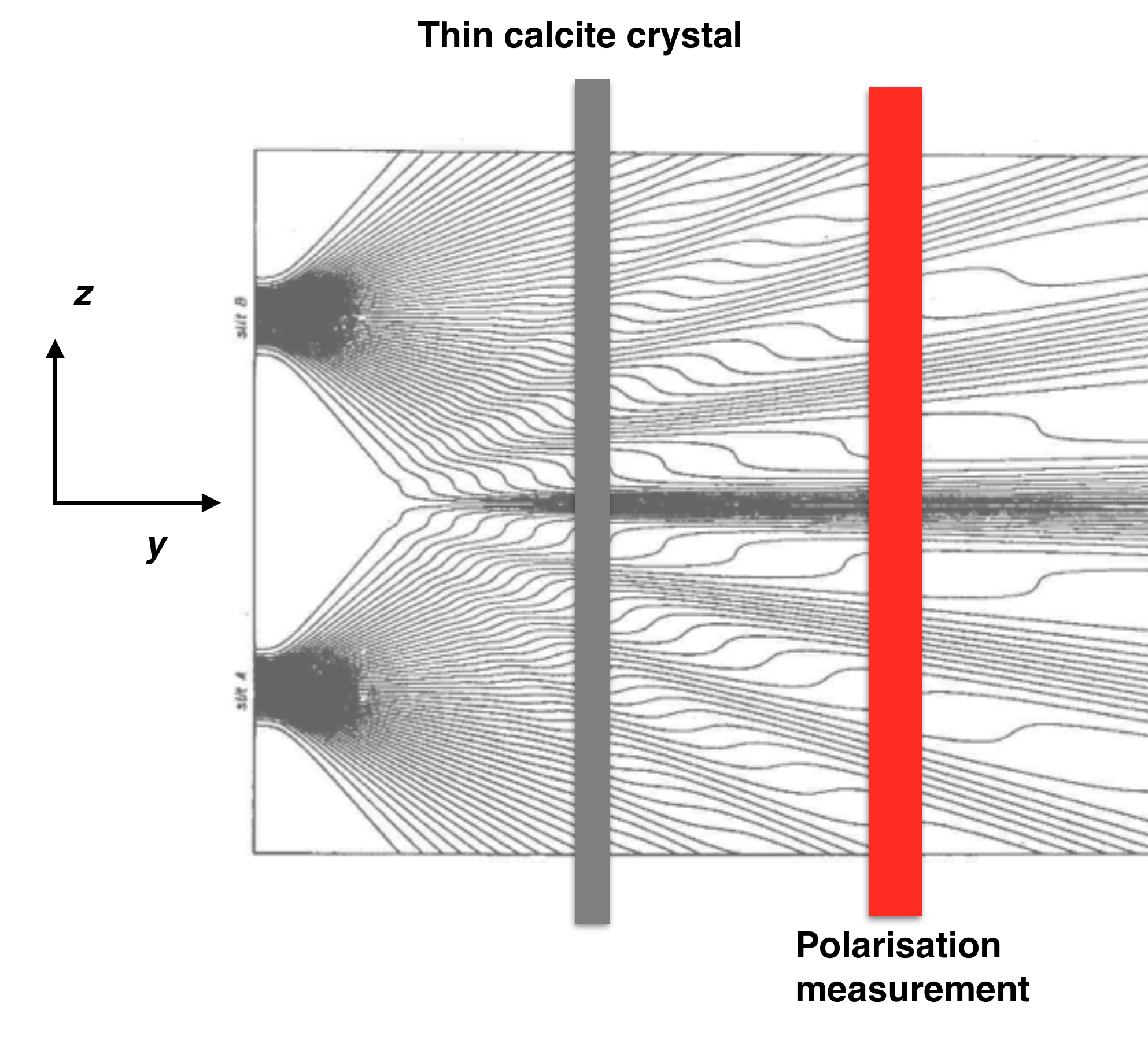} 
   \caption{Schematic Weak Measurement Apparatus for $\langle P_z\rangle_W$.}
   \label{fig:KS}
\end{figure}

In the actual experiment single photons are fed into a 50-50 beam splitter before being re-combined using a two collimated fibre couplers that act as the two slits.  The recombined beam is then passed through a thin birefringent calcite crystal, which introduces a weak effect by inducing a phase change in the overall wave function.  This phase change is proportional to $\langle P\rangle_W$ as shown in equation (\ref{eq:final state}) and can then be revealed in a strong measurement on the polarisation vector, following the principles outlined in section 2.2.

 Figure \ref{fig:KS} shows a schematic view of the apparatus required to measure $\langle P_z\rangle_W$, the transverse momentum at various points between the slit system and the final screen on which the fringes can be displayed.  The initial beam is polarised in some definite direction before being passed through the thin birefringent crystal.  The phase change introduced by the crystal leaves the spatial wave function unchanged,  but the polarisation vector is rotated by a small amount. 
 
If  the interaction Hamiltonian coupling the weak device to the beam is of the form $H_I=g(t)\hat P.\hat S$, where $\hat S$ is the polarisation operator, then the final wave function will be of the form given in equation (\ref{eq:final state}).  
To extract the information we need, the beam is passed through another polariser to separate out the left handed and right handed components and the number of photon counts per unit time in each component is determined.  From these counts the transverse momentum at a set of positions, $x$,  can be found from
\begin{eqnarray*}
\langle P_z(x)\rangle_W=G\sin^{-1}\left(\frac{N_R-N_L}{N_R+N_L}\right)
\end{eqnarray*}
where $G$ is a factor that depends on properties of the birefringent crystal.  (For full details see Kocsis{\em et al.} \cite{kbrs11}). 

To construct the energy flow lines, $P_{B_Z}(x)$, is combined with the known $P_{B_y}(x)$ to construct the tangent vector to the flow line at the point $x$.  This is repeated for a set of points between the thin crystal and the final screen.  The points are chosen so that it is possible to join up the tangent vectors to produce the flow lines. (For the details of the extrapolation method used to produce these flow lines, see the paper of Kocsis {\em et al.} \cite{kbrs11}.)

These flow lines have some resemblance to those calculated by  Philippidis {\em et al} \cite{cpcdbh79} from the Schr\"{o}dinger equation using the Bohm approach.  However direct comparison should not be expected because, in the photon case, Maxwell's equations must be satisfied which gives rise to properties that are different from those arising from the  Schr\"{o}dinger equation (see Bohm \cite{db51}.)
Notice we prefer to talk about energy flow lines because, as we have seen from the relation between $P^i_B(x,t)$ and the energy-momentum tensor components $T^{0i}(x,t)$, these weak values are momentum fluxes.  Thus in the case of light, we are actually measuring the Poynting vector of the electromagnetic field.  To interpret these flow lines as ``photon trajectories' raises some deep questions which we will discuss in the next section.

\subsection{Meaning of the Energy Flow Lines.}

As we have seen, the weak values measured in the Kocsis {\em et al.} \cite{kbrs11}  experiment enables the  Poynting vector to be calculated at a number of points.  The important feature of their experiment is that only one photon enters the apparatus at a given time so this raises a couple of questions.  ``What is the meaning of of the Poynting vector for a single photon?" and ``Can each energy flow line be regarded as a trajectory for the photon?" 

The notion of a trajectory implies that we can give meaning to a photon existing momentarily at a space-time point, but remember  the photon has a sharply defined energy and momentum.  Thus it could be argued, that because of the uncertainty principle, a photon cannot exist at a point in space-time. 

However such a conclusion is based on the assumption that all physical properties are eigenvalues of some operator or other.   But as we have already pointed out in the Bohm approach, the uncertainty principle does not apply to the beables, $\bm x$ and $\bm P_B(x)$.  The uncertainty principle only tells us is that we cannot  {\em measure} position  and momentum simultaneously using a {\em strong} measurement.  In other words, we cannot associate the particle with simultaneous eigenvalues  of the operators $\bm{ \hat X}$ and $\bm{\hat P}$.  However as we have seen   $\bm P_B$ is the real part of the weak value of the operator $\bm {\hat P}$ and therefore is not ruled out on the grounds of the uncertainty principle. As $\bm x$ is the position eigenvalue in the case we are considering, the particle does not have a strong value of $\bm {\hat P}$, but it has a weak value of the same operator given by $\langle P(x)\rangle_W$.  All of this follows from the properties of the Schr\"{o}dinger formalism.
Thus the Bohm interpretation gives simultaneous $\bm x$ and $\bm P_B(x)$ to the particle that obeys the Schr\"{o}dinger equation.  

The photon, on the other hand, is not a Schr\"{o}dinger particle and therefore we must proceed with caution.  Indeed Bohm, himself, has already pointed out that we must treat the photon in a very different way \cite{db52a}.
 The reasons for coming to this conclusion are discussed in detail in Bohm and Hiley, chapter 11, \cite{dbbh93}.   Here we will simply point out that, unlike the Schr\"{o}dinger particle with its finite rest mass, the photon has no position state $|x\rangle$.  Or as Muthukrishnan, Scully and Zubairy \cite{msz03} puts it, there is no particle creation operator that creates a photon at an exact point in space. 
 
 An even simpler way of looking at it is to realise there is no way a photon can be `slowed down' to reveal a non-relativistic, classical behaviour as is the case for a Schr\"{o}dinger particle.  In the latter, we simply have to find situations where the quantum potential is negligible to find the resulting behaviour is  classical \cite{bham95}.  

In the classical limit of quantum electrodynamics,  we are left simply with a continuous field, which satisfies Maxwell's equations.  This implies we should start with the electromagnetic field and develop a way of describing a quantised field by treating the classical  field  as a `beable',  and treating the fields in an analogous way to the way we treat the  position and momentum of the Schr\"{o}dinger particle.

Following up that line of reasoning, it might be tempting  to take the $\bm E$ and $\bm B$ as the beables and then consider the energy  density $T^{00}(x)=(E^2(x)+B^2(x))/2$  and Poynting vector $T^{0j}(x)=\bm E(x)\times \bm B(x)$ as giving information about the local energy and momentum.  Recall the Poynting vector give us the momentum of an electromagnetic beam 
\begin{eqnarray*}
\bm p=\frac{1}{4\pi c}\int(\bm{E\times B})d\tau
\end{eqnarray*}
Clearly we can regard this momentum as being shared between the individual photons in the beam.  Indeed dividing the Poynting vector by the number of photons in the beam gives us the average momentum carried by a photon.  The crucial question as far as the experiment of Kocsis {\em et al} is concerned is whether we can use this momentum to calculate the `trajectory' of a photon. 

Unfortunately in this relativistic case we do not get a consistent set of trajectories in all frames.  This difficulty has already been discussed in Bohm, Hiley and Kaloyerou~\cite{dbbhpk87} so we will not repeat the arguments here.  Instead lets us give an example to illustrate the type of problem that arises.   Consider an electrostatic field $\bm E$ uniform in the $x$-direction.  The Poynting vector is zero, and this would correspond to the photons at rest.  Consider a Lorentz transformation, again in the $x$-direction.  $\bm E$ is unchanged and there is still no $\bm B$ field, so the Poynting vector remains zero and the photons are still at rest, but clearly the photons cannot be at rest in both frames.  

These and other arguments show that we cannot obtain a consistent way of attaching the results obtained from these fields to construct meaningful photon trajectories.  This criticism about trajectories is in no way a criticism of the energy flow results obtained in the experiment of Kocsis {\em et al.} \cite{kbrs11}. The experiment is still important because it demonstrates for the first time how weak measurements can be used to measure components of the energy-momentum tensor in the quantum domain.

\section{Weak Measurements with Atomic Beams.}

The objections against photon trajectories do not apply to  Schr\"{o}dinger particles.  In the Bohm approach  the $T^{0\mu}$ components are given a meaning in terms of the quantum Hamilton-Jacobi equation where they appear as Bohm momentum $\bm {P_B}$ and the Bohm energy $E=-\partial S/\partial t$.  The quantum Hamilton-Jacobi equation (\ref{eq:HJ}), like its classical counterpart, is simply an expression for energy conservation.  We know that the classical Hamilton-Jacobi gives an ensemble of possible trajectories.  The presence of $Q$ in equation (\ref{eq:HJ}) does not deny the possibility of an ensemble of trajectories, it merely alters the form of the trajectories as shown in Figure \ref{fig:traj1}.

Furthermore if $Q$ becomes negligible then the resulting trajectories become the classical trajectories as was shown in Hiley and Mufti \cite{bham95}.  Since we know in the classical limit, a particle follows one of these trajectories and as these trajectories deform into their corresponding quantum counterparts, it is natural to assume that in the quantum case of a Schr\"{o}dinger particle, we can retain the notion of a localised particle in this case.  Any response of this particle to the global environment is `communicated' through the quantum potential $Q$.  

To calculate these trajectories in a given experiment, say a two-slit interferometer,  we simply integrate the expression for $\bm P_B(x)=m\;d\bm r_B/dt.$ which gives us the set of trajectories shown in Figure \ref{fig:traj1}.  In order to establish the experimental legitimacy of these trajectories we propose to carry out an experiment along the lines of Kocsis {\em et al.} \cite{kbrs11}, but in this case using atoms.  The atoms must be neutral but possess a magnetic moment so that we can use the interaction Hamiltonian given in equation (\ref{eq:HI}). This means we must replace the thin birefringent crystal with some form of uniform electric field.  This will ensure the momentum operator is coupled to the spin of the atom in an appropriate way.  

We are at the moment in the planning stage of developing a two-slit interference experiment using a  beam of atoms.  In this experiment we plan to measure the weak value $\langle P_z\rangle_W$ which will allow us to calculate the transverse Bohm momentum.  This value, together with the momentum in the direction of the beam, which in this case is along the $y$-axis (see Figure \ref{fig:KS}) will allow us to determine tangent momentum vectors at a series of points in the interference region.  From these an ensemble of atomic trajectories can  be calculated along the lines indicated in the experiment of Kocsis {\em et al.} \cite{kbrs11}.  Thus we will be able to compare the experimental results with those predicted by the Bohm approach.

\section{Conclusion.}

Following on from the important experimental work of Kocsis {\em et al.} \cite{kbrs11}, who measured the energy flow lines shown in Figure \ref{fig:emflow}, we have, in this paper, discussed the possibility of using  similar techniques to measure the two-slit trajectories of atoms.  The advantage of our proposed experiment is that in the case of atoms, a well defined  notion of a trajectory is provided by the Bohm approach to the quantum formalism.  This approach uses the  real part of Schr\"{o}dinger equation to define what we mean by a trajectory.  Thus we have a much clearer meaning of the notion a trajectory for an atom than can be given to  a photon.  In the case of the atom, its classical behaviour is described by the Hamilton-Jacobi equation, which is clearly the limit of equation~(\ref{eq:HJ}), whereas in the case of the photon, the classical behaviour is  that of a field.

Since the Bohm approach based on the Schr\"{o}dinger particle predicts well defined trajectories which have actually been calculated by Philippidis {\em et al.} \cite{cpcdbh79} (See Figure~\ref{fig:traj1}), our proposals will show whether the trajectory calculations agree with experiment.

\section{Acknowledgements}
The authors would like to thank  Marcus Arndt of the University of Vienna, Peter Barker of UCL, and Jim Clark of the Cockcroft Institute for their help and advice in the design stages of the proposed atomic experiment.

\vspace{2cm}

\section{Appendix}

As we have already indicated, it is not possible to treat photons in the same way as Schr\"{o}dinger particles in the Bohm approach.  Since in the classical limit we have to deal with fields, we  choose the field and its conjugate momentum to be  the beables of the theory \cite{db52a,dbbhpk87, dbbh93}.  Since we are dealing with fields, the key question then is to ask how the concept of a photon appears in this approach.

We start with a Fourier analysis of the vector potential, namely,
\begin{eqnarray*}
\bm A(\bm x)=(4\pi/V)^{1/2}\sum_{k,\mu}\epsilon_{k,\mu}q_{k,\mu}e^{i \bm {k.x}}
\end{eqnarray*}
with $q_{k,\mu}^*=q_{k,\mu}$.  We are working in the gauge $\bm {\nabla.A}=0$ and restricting our consideration to the transverse parts on the electromagnetic field.

Introducing the conjugate momentum $\bm \Pi_{k,\mu}=\partial q_{k,\mu}^*/\partial t$.  Then the transverse part of the electric field is
\begin{eqnarray*}
\bm E(\bm x)=-\frac{1}{c}\frac{\partial \bm A(\bm x)}{\partial t}=-\left(\frac{4\pi}{Vc^2}\right)^{1/2}\sum_{k.\mu}\epsilon_{k,\mu}\bm \Pi_{k,\mu}^*e^{i\bm{k.x}}.
\end{eqnarray*}
and the transverse part of the magnetic field is
\begin{eqnarray*}
\bm B(\bm x)=\nabla\times \bm A=-(4\pi/V)^{1/2}\sum_{k,\mu}(\bm k\times\bm\epsilon_{k,\mu})q_{k,\mu}e^{i\bm{k.x}}.
\end{eqnarray*}
The beables of the theory are then taken to be $q_{k, \mu}$ and $\bm \Pi_{k,\mu}$  The quantum Hamilton-Jacobi equation equation becomes
\begin{eqnarray*}
\frac{\partial s}{\partial t}+\sum_{k,\mu}\frac{\partial s}{\partial q_{k,\mu}} \frac{\partial s}{\partial q_{k,\mu}^*}+\sum_{k.\mu}(kc)^2q_{k,\mu}q_{k,\mu}^*\hspace{3cm}\\
-\frac{\hbar^2}{2R}\sum_{k,\mu}\frac{\partial^2 R(\dots q_{k,\mu}\dots)}{q_{k,\mu}q_{k,\mu}^*}=0
\end{eqnarray*}
The equation of motion of the $q_{k,\mu}$ derived fro the Hamiltonian
\begin{eqnarray*}
H=\sum_{k,\mu}(\bm \Pi_{k,\mu}\bm\Pi_{k,\mu}^*+k^2c^2q_{k,\mu}q_{k,\mu}^*)
\end{eqnarray*}
becomes
\begin{eqnarray*}
\ddot q_{k,\mu}+k^2c^2q_{k,\mu}=\frac{\partial}{\partial q_{k,\mu}^*}\left(\frac{\hbar^2}{2R}\sum_{k',\mu'}\frac{\partial^2}{q_{k',\mu'}q_{k',\mu'}^*}\right).
\end{eqnarray*}
Since Maxwell's equations for empty space follow when the right-hand term is zero, we see that the presence of this term profoundly modifies the behaviour of the electromagnetic field.  This modification means that the oscillator, $q_{k,\mu}$ can transfer a large quantity of energy and momentum even when $q_{k,\mu}$ is very small, because when $q_{k,\mu}$ is small the right-hand term may become very large.  It is this rapid non-local and non-linear term that enables us to explain the notion of a photon.

In practice we cannot produce a beam of photons with all the photons having a sharp $\delta$-function distribution in momentum. Usually there is a small distribution in momenta $f(\bm k-\bm k_0)$ so that the wave functional takes the form
\begin{eqnarray*}
\Psi=\sum_{k,\mu}f_\mu(\bm k-\bm k_0)q_{k,\mu}e^{-kct}\Psi_0
\end{eqnarray*}
where $\Psi_0$ is the ground state of the electromagnetic field which is given by
\begin{eqnarray}
\Psi_0=\exp[-\sum_{k,\mu}(kcq_{k,\mu}q_{k,\mu}^*+ikct/2)].	\label{eq:GS}
\end{eqnarray}
  What Bohm showed was that the excess energy over the ground state is only appreciable within a region in which the wave packet $g(\bm x)$ is appreciable,  where
\begin{eqnarray}
g(\bm x)=\sum_{k,\mu}f_\mu(\bm k-\bm k_0)e^{i\bm{k.x}}\epsilon_{k,\mu}.\label{eq:wavepacket}
\end{eqnarray}

In order to understand the photoelectric effect, we must introduce the interaction Hamiltonian
\begin{eqnarray*}
H=\frac{1}{2m}[\bm p-(e/c)\bm A(\bm x)]^2.
\end{eqnarray*}
The photoelectric effect corresponds to the transition of a radiation oscillator from an excited state to the ground state, while the atomic electron is ejected with an energy $E=h\nu-I$, where $I$ is the ionisation potential of the atom. The initial wave functional of atom plus the field  containing only one quantum is
\begin{eqnarray*}
\Psi_i=\psi_0\exp(-iE_0t/\hbar)\sum_{k,\mu}f_\mu(\bm k-\bm k_0)q_{k,\mu}e^{-kct}\Psi_0(\dots q_{k,\mu}\dots).
\end{eqnarray*}
Now solving the Schr\"{o}dinger equation, we obtain an asymptotic wave functional
\begin{eqnarray*}
\Psi_f=\Psi_0(\dots q_{k,\mu}\dots)\sum_{k,\mu}f_\mu(\bm k-\bm k_0)\hspace{3cm}\\
\times\frac{\exp[i\bm{k'.r}-i\hbar(k'^2/2m)t]}{r}g_\mu(\theta,\phi, k'),
\end{eqnarray*}
Where the energy of the outgoing electron is $E=\hbar^2k'^2/2m=\hbar kc+E_0$  The function $g_\mu(\theta,\phi, k')$ is the amplitude associated with the $\psi$-field of the outgoing electron.  The outgoing electron wave packet, centred on $r=(\hbar k'/m)t$ will eventually become separated from its initial wave function $\psi_0$.   If the electron happens to enter the outgoing packet, the initial wave function can subsequently be ignored. The system then acts for all practical purposes as if its wave field were given by equation (\ref{eq:GS}) from which we conclude  that the radiation field is in the ground state, while the electron has been liberated.  It is readily shown that, as in the usual interpretation, the probability that the electron appears in the direction $\theta, \phi$ can be calculated from $|g_\mu(\theta,\phi, k')|^2$.  Thus the energy swept from the quantum field is just equivalent to one photon energy.  

The picture is that the one quanta of energy is spread over the extent of the wave packet described by equation (\ref{eq:wavepacket}).  Then during the interaction between the field and the electron, one quantum of  energy is swept from the field and this is used to release the electron.  Thus in this picture, the photon does not exist at a point like a Schr\"{o}dinger particle.  Bohm shows that the same picture can be made to explain the Compton effect.  In fact other examples exhibiting this behaviour will be found in  Bohm, Hiley and  Kaloyerou \cite{dbbhpk87} and Kaloyerou \cite{pk94}.

%References 1

\end{document}